\documentclass[12pt,preprint]{aastex}
\def\labelitemi{$\circ$}
\pdfoutput=1

\slugcomment{{\it NASA LAW}, October 25-28, 2010, Gatlinburg }

\begin{document}

\title{Cosmology and Fundamental Physics and their Laboratory Astrophysics Connections}

% Authors and affiliation #1
\author{W. C. Haxton}
\affil{Department of Physics, MC 7300, University of California, Berkeley, CA 94720-7300 and \\
Lawrence Berkeley National Laboratory, MS 70R0319, Berkeley, CA 94720-8169}
\email{haxton@berkeley.edu}

% ABSTRACT SECTION STARTS HERE
\begin{abstract}

The  Decadal Survey of Astronomy and
Astrophysics created five panels to identify the science themes that would define the 
field's research frontiers in the coming decade.  I will describe the conclusions of
one of these, the Panel on Cosmology and Fundamental Physics, and comment
on their relevance to the discussions at this meeting of the NASA Laboratory
Astrophysics community.
   
\end{abstract}

%  INTRODUCTION STARTS HERE:
\section{Introduction}

The \cite{panel} convened five panels to
consider the science themes that would define the field's research frontiers in the next decade.
One of these, the Panel on Cosmology and Astrophysics, had a particularly broad 
mandate that included topics of interdisciplinary interest.  The panel was chaired by
David Spergel and included David Weinberg (vice chair), Rachel Bean, Neil Cornish,
Jonathan Feng, Alex Filippenko, Marc Kamionkowski, Lisa Randall, Eun-Suk Seo, 
David Tytler, Cliff Will, and myself.  The organizers of this Laboratory Astrophysics Workshop have asked me to
summarize the Panel's conclusions and comment on their relevance to laboratory
astrophysics and future NASA missions.

The context for Panel discussions was established by a set of
recent discoveries that have strengthened the links between astrophysics/cosmology 
and fundamental physics conducted in terrestrial laboratories.
These include
\begin{list}{\labelitemi}{\leftmargin=0.25cm}
\item The development of a relatively simple cosmological model fitting astronomical data,
Lambda Cold Dark Matter, with parameters known to better than 10\% and with immediate
implications for beyond-the-standard-model physics.
\item Cosmic microwave background (CMB) and large-scale structure (LSS) studies that appear
consistent with the predictions of inflation: a nearly flat universe with a matter distribution
that is Gaussian with nearly scale-invariant initial fluctuations.
\item CMB confirmation of the Big Bang nucleosynthesis (BBN) conclusion that baryons
comprise about 4\% of the closure density $\Omega_c$, so that dark matter must be primarily nonbaryonic.
\item Supernova data indicating that the expansion of the universe is accelerating, consistent
with dark energy dominance of the universe's present energy density.
\item Astrophysical $\nu$ discoveries, from the Sun and from cosmic rays (CRs) impinging on Earth,
that show neutrinos have mass and undergo flavor oscillations, providing the first
direct evidence of physics beyond the standard model (and the first identification of a
component of the dark matter).
\item The identification of a cutoff in ultra-high-energy (UHE) CRs consistent with the
expected GZK scattering off the CMB.  Thus the universe may be opaque to us at
cosmological distances and asymptotic energies, apart from UHE $\nu$s.
\end{list}

The Panel considered community input, generally provided as white papers,
on a wide range of topics: the early universe; the
CMB; probes of LSS through observations of galaxies, intergalactic gas, or their associated
gravitational distortions; determinations of cosmological parameters; dark matter; dark energy;
tests of gravity; astrophysical measurements of physical constants; and the fundamental
physics that might be derived from astronomical messengers ($\nu$s, $\gamma$s,
CRs).  Among the white papers considered, several addressed either laboratory
astrophysics or theory and computation.

The Panel's response was formulated around four ``big questions:"
1)  How did the universe begin  (the mechanism behind inflation)?
2) Why is the universe accelerating (the nature of the dark energy)?
3) What is dark matter?
4) What are the properties of neutrinos?
Gravitational wave astronomy was designated as the discovery area.

This meeting's organizers have asked me to summarize the Panel's conclusions,
commenting on their connections to laboratory astrophysics and NASA missions.
In this written version of my talk I will focus the last two of the four questions, in part
because I know these areas best, but also because they may have substantial
connections to laboratory astrophysics.
Here ``laboratory astrophysics" is defined quite broadly, given that the Panel's
charge included the intersection of astrophysics and astronomy with the 
particle and nuclear physics programs of major accelerator facilities,
and with a broad array of ground-based detectors for dark matter,
$\nu$s, CRs, and related studies.

\section {What is Dark Matter?}
The majority of matter in the universe is dark, invisible to us apart from its
gravitational effects on the structure we do see.   In addition to its deep roots in
cosmology and astrophysics, the dark matter (DM) problem is central to
high-energy physics, where DM particles may be discovered in the debris
from collisions between ordinary particles, and in underground science, where
the recoil of detector nuclei may indicate interactions with dark matter particles.
(For a recent review of the topics summarized here, see \cite{feng}.)

DM was first postulated to account for the anomalous velocity rotation
curves of  galaxies.  DM particles must be stable or long-lived, cold
or warm (sufficiently slow that they can seed structure formation), gravitationally
active, but without strong couplings to themselves or to baryons.  The DM/dark energy
contributions to the universe's total energy density evolves with redshift, with
the former dominant early and the latter dominant today.  Two leading DM candidates
are Weakly Interacting Massive Particles (WIMPs) and axions.  WIMPS are
intriguing because the properties necessary for astrophysics
match expectations that new particles will be found at the mass generation 
scale of the standard model of 10 GeV - 10 TeV.  The WIMP ``miracle" is the
observation that the annihilation cross section for massive, weakly interacting
particles natural leads to the expectation that $\Omega_\mathrm{WIMP} \sim 0.1.$

\begin{figure}[ht]
\centering
\includegraphics[angle=0,scale=.50]{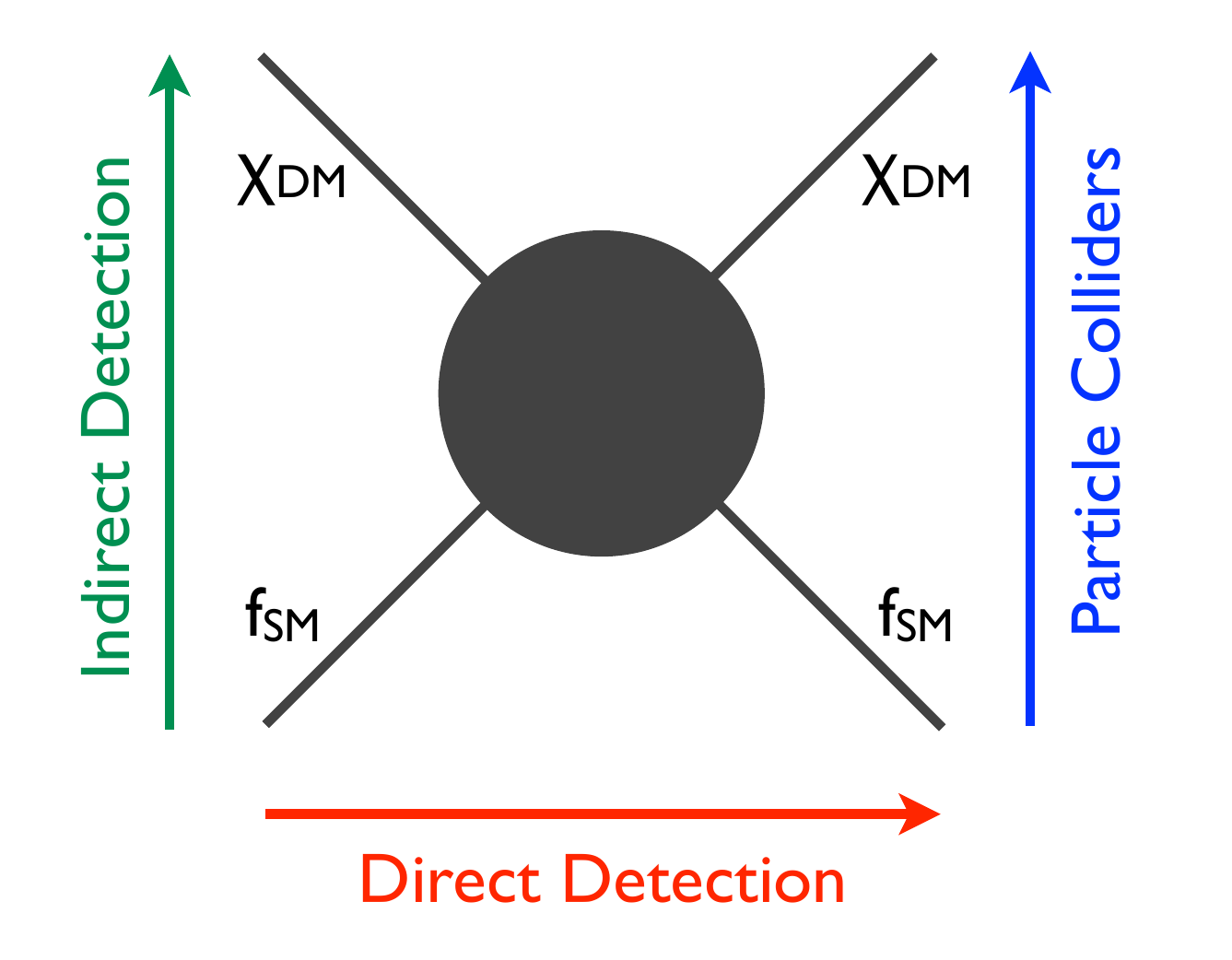}
\caption{Three strategies for detecting dark matter particles $\chi_\mathrm{DM}$ via
their interactions with standard-model particles f$_\mathrm{SM}$.\label{fig:Wimp}}
\end{figure}

Figure \ref{fig:Wimp} illustrates three avenues of attack on the DM problem:
direct detection where a DM particle $\chi_\mathrm{DM}$ scatters off a standard-model particle f$_\mathrm{SM}$, causing
recoil; indirect detection, where DM particles annihilate or decay
into ordinary particles that can then be detected; and particle collider experiments,
where DM particles are produced from the scattering of ordinary particles and identified
from the missing energy.
DM particle interactions can be either independent (SI) or dependent (SD) on target spin, depending on
parameter choices in the underlying model, and while their predicted cross sections span a wide range,
$\sigma_{SI} \sim 10^{-45}$ cm$^2$ is a representative value.  Current detectors
in the $\sim$ 10-100 kg range are probing DM particle-nucleon cross sections well
below $10^{-43}$ cm$^2$ for DM particle masses
of $\sim$ 100 GeV.  The international program is focused on developing new detectors
in the 1-10 ton range, using media such as ultra-clean nobel-gas liquids, with sensitivities
of a few events per year, or $\sigma_{SI} \sim 10^{-47}$ cm$^2$.  There have been
claims of detection, interpreted as low-mass WIMPs, but no consensus has been reached.

This field has a number of laboratory and theory needs, including:
\begin{list}{\labelitemi}{\leftmargin=0.25cm}
\item support for direct searches, including detector R\&D and the development of deep underground locations for detectors, 
as energetic neutrons produced by penetrating CR muons are an important background;
\item clean-room facilities to control environmental activities, as trace radionuclides within detectors or in surrounding
materials are a second major background source;
\item nuclear theory for estimates of WIMP SI form factors and SD cross sections;
\item for direct production experiments, facilities like the LHC that can reach the energies
necessary for $\chi_\mathrm{DM}$ creation; and
\item  for indirect detection searches, astrophysical modeling that will allow observers to
distinguish WIMP annihilation signals from other high-energy astrophysics phenomena.
\end{list}
Ideally, multiple lines of investigation will lead to DM detection.  An
attractive scenario is the discovery of supersymmetry at the LHC and the identification
of a lightest stable SUSY particle; direct detection of cosmic WIMPS with consistent properties;
and consequently, improved constraints on the local DM density and its effects on structure at subgalactic
scales, testing the paradigm of cold, collisionless, stable DM.

\section{What are the Properties of Neutrinos?}
\begin{figure}[ht]
\centering
\includegraphics[angle=0,scale=.50]{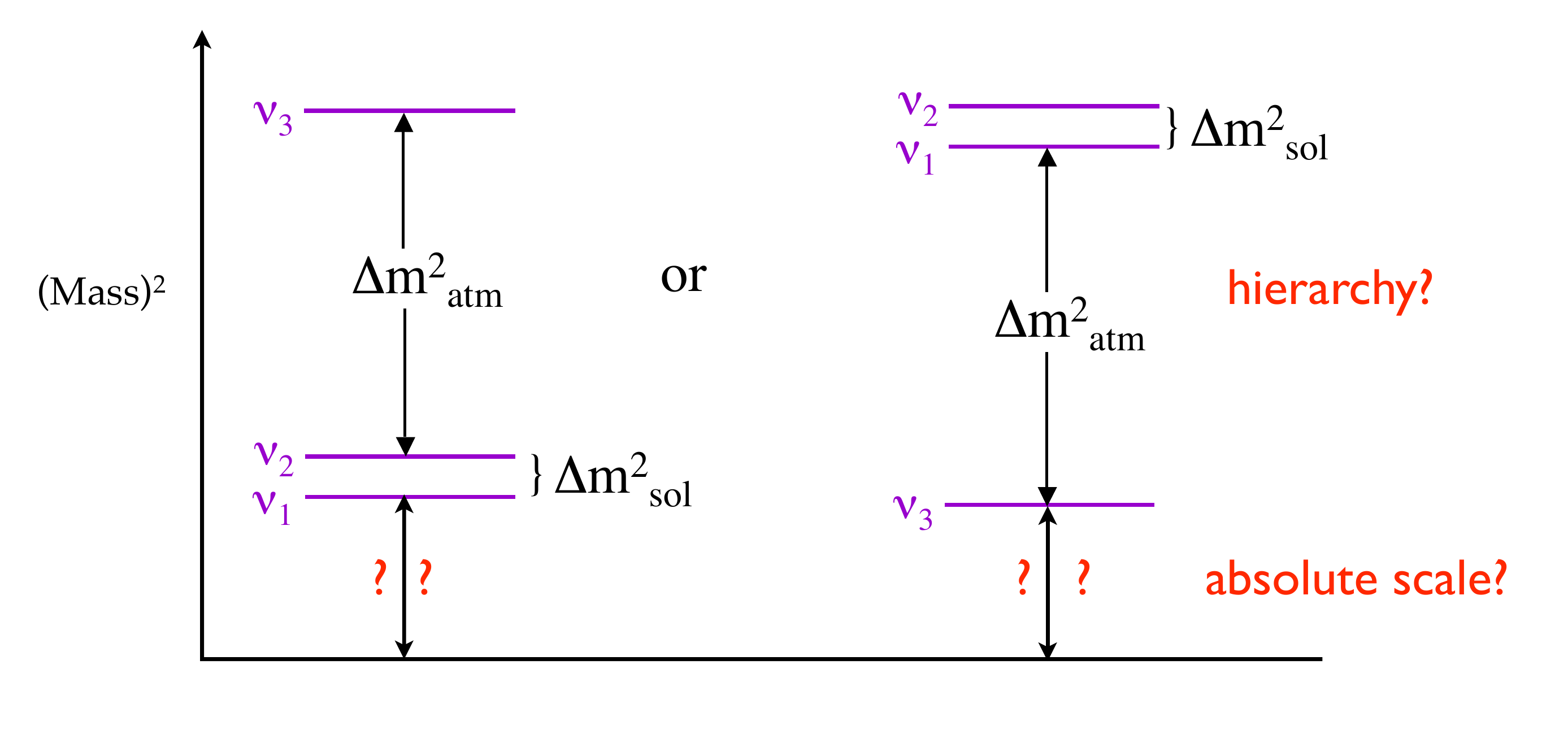}
\caption{The normal (left) and inverted (right) $\nu$ mass hierarchies are
both consistent with existing oscillation data.\label{fig:hierarchy}}
\end{figure}

[For a review of topics discussed in this section, see the \cite{APS}.]
 Neutrino astrophysics has rich intersections with laboratory nuclear and particle astrophysics.
A 1958 measurement showing that the cross section for $^3$He+$^4$He is
a 1000 times larger than then expected was crucial to the first solar $\nu$ experiment:
this reaction leads to pp chain branches producing
higher energy $^7$Be and $^8$B $\nu$s, which the Cl experiment could detect.
The discrepancies that emerged from the Cl experiment stimulated 30 years of careful laboratory cross
section measurements and the development of a standard solar model (SSM)
capable of predicting the Sun's core temperature to $\sim$ 1\%.  When the pattern of solar
$\nu$ fluxes did not match that predicted by the model nor by any plausible variation in that
model, expectations grew that a more fundamental problem must exist.   Finally the atmospheric 
$\nu$ measurements of Super-Kamiokande and the solar $\nu$ measurements made by SNO 
demonstrated that neutrinos have mass and that two distinct oscillations occur, governed by the
mass splittings $\Delta m_\mathrm{sol}^2$ and $\Delta m_\mathrm{atm}^2$ (see Fig. \ref{fig:hierarchy}).

Unsettled issues in $\nu$ physics with important implications for astrophysics include
the absolute scale of neutrino mass;
the origin of matter in the cosmos;
$\nu$ properties affecting energy transport and nucleosynthesis in extreme astronomical environments; and
the high-energy limits of astrophysical accelerators.
The associated laboratory astrophysics is very ``high end," predicated on next generation $\nu$
experiments requiring new beamlines and massive underground detectors.  

The $\nu$ differs from other fermions of the standard model because it has no charge or 
other additively conserved quantum number.  Consequently it can have two
kinds of mass terms, Dirac and Majorana, while other particles must be Dirac fermions. This 
provides a natural explanation for an otherwise mysterious fact, that $\nu$s are much less massive 
than other standard-model particles.  The $\nu$ mass matrix can be written schematically as
\begin{equation}
\left( \begin{array}{cc} M_L \sim 0 & M_D \\ M_D^\dagger & M_R \end{array} \right) \Rightarrow 
m_\nu^\mathrm{light} \sim M_D \left( {M_D \over M_R} \right),
\end{equation}
yielding a light $\nu$ mass that is proportional to the Dirac mass, multiplied by $M_D/M_R$, the 
ratio of the Dirac mass to the right-handed Majorana mass.   If $M_R \gg M_D$, one has a natural
explanation for the smallness of the $\nu$ mass, relative to the Dirac mass $M_D$ of other fermions.
Indeed, based on what we have learned from $\nu$ oscillations, the necessary $M_R \sim 0.5 \cdot 10^{15}$ GeV,
close to the $\sim 10^{16}$ GeV grand unified mass scale.   Thus $\nu$s not only involve an entirely
distinct mass generation mechanism, but that mechanism could depend on UHE physics that otherwise is far beyond
experimental reach.   There is great interest in determining the absolute $\nu$ mass and whether
Majorana masses exist.  (The latter question involves the process of neutrinoless
$\beta \beta$ decay, which I will not discuss here.)

Neutrino oscillation mass differences tell us that the $\nu$ mass could be as small of 0.05 eV.  The
most promising test we have of such small $\nu$ masses is cosmology: LSS is influenced by $\nu$s,
because they are relativistic and free-streaming, and thus suppress the growth of LSS.  The smaller
the mass, the longer they remain relativistic, and the larger the scale at which they suppress the growth of
structure. The critical wave number is related to the mass by
\begin{equation}
k_\mathrm{free~streaming} \sim 0.004 \sqrt{m_\nu/0.05\mathrm{~eV}}~\mathrm{Mpc}^{-1}.
\end{equation}
Neutrinos are a unique DM component because they transition from relativistic to nonrelativistic matter
with expansion.  Thus their effects are both scale and red-shift dependent.  To see effects due to $m_\nu \sim$ 0.05 eV,
a sensitivity to DM at 0.1\% of $\Omega_c$ must be achieved.  (However, their effects in suppressing power at
large scales and low Z can be several per cent.)   The current best limits correspond to
1.3\% of $\Omega_c$.  Future analyses that
combine large data sets with different sensitivities to scale and redshift will clearly have the most impact
on improving sensitivity to $\nu$ mass -- assuming
systematic uncertainties of multiple data sets are under control.  As the sensitivity of LSS surveys typically scales as
$\sqrt{N}$, where N is the number of modes, one needs new LLS surveys
that exceed past ones by a factor $\sim$ 100.  The Panel discussed a variety of planned surveys
of high-redshift galaxies, QSOs, and the CMB that might achieve such improvements.
Some combination of such surveys,  data from 21-cm radio telescopes with large collecting areas ($\sim$ 0.1 km), and weak lensing studies
could reach the sensitivity to detect $\nu$ mass at the 0.05 eV lower bound.

Other open $\nu$ questions affect astrophysical phenomena such as Type II supernova explosions.  Figure
\ref{fig:hierarchy} illustrates an ambiguity in the $\nu$ mass pattern: both the normal and inverted
hierarchies are compatible with existing oscillation data.   In solar $\nu$ measurements, the MSW
effect alters oscillation results: the $\nu_e$ becomes heavier in matter, generating a level crossing
at some critical density where the effective mass generated by $\nu_e$ interactions with solar matter
just cancels the vacuum mass difference.
The imprint of matter effects on the solar $\nu$ spectrum is how we know
the sign of $\Delta m_\mathrm{sol}$.  But for the atmospheric $\nu$ oscillations, $\Delta m_\mathrm{atm}$ is 
too large for terrestrial matter effects to enter.  Consequently, we do not know the sign of $\Delta m_\mathrm{atm}$,
leading to the ambiguity illustrated in Fig. \ref{fig:hierarchy}.

However, in core collapse supernovae, $\nu$s decouple from the matter at very high density, $\sim 10^{11}$ g/cm$^3$.
As they free-stream through the carbon zone, at a density of about 10$^4$ g/cm$^3$, they encounter the critical
density where a crossing occurs for $\Delta m_\mathrm{atm}$.  For the normal hierarchy, the crossing is between 
the $\nu_e$ and its heavy-flavor counterpart.   But for the inverted hierarchy, the crossing is between the
corresponding $\bar{\nu}$s.  That is, the sign of the density-dependent effective mass is opposite for $\nu_e$s
(which become heavier in matter) and $\bar{\nu}_e$s (which become lighter).  As the $\nu_e$s and $\bar{\nu}_e$s
from a supernova are expected to be less energetic that the heavy-flavor $\nu_x$s, the flavor swap will lead
to either anomalously hot $\nu_e$s or anomalously hot $\bar{\nu}_e$s -- depending on the hierarchy.

This crossing also depends on a third mixing angle $\theta_{13}$ that is currently unknown, bounded only by reactor
$\bar{\nu}_e$ disappearance experiments.  The phenomena described above require $\theta_{13} \gtrsim 10^{-4}$.   As
next-generation terrestrial $\nu$ experiments have a more modest sensitivity goal of $\theta_{13} \gtrsim 10^{-2}$, it is quite
possible that high precision measurements of the $\nu$  flux and flavors from the next galactic supernova could also
be important in probing $\theta_{13}$.

This issue is important to another cosmological puzzle, the origin of matter (baryons) in the universe.   We know
some process broke the symmetry between matter and antimatter in the Big Bang: an excess of matter over
antimatter led to incomplete annihilation, so that today baryons comprise 4\% of $\Omega_c$.
One of several requirements for baryogenesis in the early universe is CP violation -- yet the known
CP violation in the standard model is too weak to drive this process.   Thus there likely is some undiscovered source of CP violation.
One consequence of nonzero $\nu$ masses is that a Dirac CP phase $\delta$ now appears in the $\nu$ mass matrix.  This
phase could be responsible for the generation of baryons through the process of leptogenesis: the CP violation originates
with $\nu$s, but is communicated to the baryons through interactions.
Given what we know about other mixing angles, the condition for CP violation to be large among the $\nu$s (in
contrast to quarks, where other small mixing angles suppress the effects
of CP violation) is that neither $\delta$ nor $\theta_{13}$ is small.  Thus, if we
detect the effects of $\theta_{13}$ on supernova $\nu$ oscillations, this would be a major step forward, ensuring that
$\theta_{13} \gtrsim 10^{-4}$.

\section{Conclusion}
 The laboratory astrophysics $\leftrightarrow$ astrophysics/cosmology intersection described in this talk is sometimes
 conventional -- e.g., the nuclear cross section measurements done in support of solar $\nu$ experiments or BBN
 modeling --
 but more often unconventional.   For example, in determining the overall scale of neutrino mass, the experiments 
 done in the laboratory (tritium $\beta$ decay, $\beta \beta$ decay, reactor
 and accelerator $\nu$ oscillation searches) complement what can be done in cosmology.
 Each probes an important quantity -- tritium $\beta$ decay probes the masses of $\nu$s in proportion to their
 coupling probability to the electron, $\beta \beta$ decay probes the Majorana mass, 
 oscillations test $m_\nu^2$ differences, and cosmology responds to
 the sum of the $\nu$ masses -- but the quantities are different.  So the traditional relationship has given way to one where
 the universe has be viewed as another laboratory, one that is playing a very important role in pushing back the
 frontiers of precision particle physics.  This is a theme that may have been best expressed in another NRC study,
 From Quarks to the Cosmos.
 
 In the coming decade the community hopes to build new $\nu$ beamlines and underground detectors on the 
 $10^2-10^3$ kiloton scale to determine the hierarchy through matter effects, measure $\nu$ CP violation, and fix
 all mixing angles and mass differences to high precision.   Cosmology could be an equal partner in this effort:
 if the effects of $\nu$ mass on LLS can be determined to sufficient accuracy, both the absolute scale and hierarchy
 questions might be resolved in this way.  
 
 In DM the situation is similar.   One of the goals of the LHC is to find the new particles that are expected to accompany
 TeV-scale physics.  We also have a new generation of massive DM detectors that will be mounted on Earth
 (or more precisely, within the Earth) that will be probing the cosmological flux of DM particles.   One hopes that 
 both endeavors succeed, and that we will be able to reconcile the cosmological properties of DM with those 
 determined from collider experiments.

\acknowledgments
I thank the organizers for the opportunity to attend this workshop, and my colleagues on the Cosmology and
Fundamental Physics Panel for sharing their perspectives on this field.  This work was supported
by the US Department of Energy, Office of Nuclear Physics.

\end{document}